\documentclass[10pt,aps,prd,nofootinbib,superscriptaddress,twocolumn,preprintnumbers,balancelastpage]{revtex4}

\usepackage{graphicx,epsfig,psfrag,amssymb,hyperref}
\usepackage{multirow}
\usepackage{color,graphicx,epsfig,psfrag,amsmath,empheq}
\usepackage{bm}
\usepackage{mathrsfs,amsfonts,color}
\usepackage{slashed}
\usepackage{hepunits}
\usepackage{color}
\usepackage{enumitem}

\newcommand{\be}{\begin{equation}}
\newcommand{\ee}{\end{equation}}

\newcommand{\E}{\mathbf{E}}

\newcommand{\B}{\mathbf{B}}
\newcommand{\Pol}{\mathbf{P}}
\newcommand{\M}{\mathbf{M}}
\newcommand{\J}{\mathbf{J}}
\newcommand{\g}{g_{a \gamma \gamma}}

\newcommand{\x}{\mathbf{x}}

\newcommand{\SNR}{{\rm SNR}}

\begin{document}

\title{Probing ALPs and the Axiverse with Superconducting Radiofrequency Cavities}


\author{Zachary Bogorad}
\email{zbogorad@mit.edu}
\affiliation{Laboratory for Nuclear Science, Massachusetts Institute of Technology, Cambridge, MA 02139, U.S.A.}

\author{Anson Hook}
\email{hook@umd.edu}
\affiliation{Maryland Center for Fundamental Physics, Department of Physics, University of Maryland, College Park, MD 20742, U.S.A.}

\author{Yonatan Kahn}
\email{ykahn@uchicago.edu}
\affiliation{Kavli Institute for Cosmological Physics, University of Chicago, Chicago, IL 60637, U.S.A.}
\affiliation{University of Illinois Urbana-Champaign, Urbana, IL 61801, U.S.A.}

\author{Yotam Soreq}
\email{yotam.soreq@cern.ch}
\affiliation{Theoretical Physics Department, CERN, CH-1211 Geneva 23, Switzerland}
\affiliation{Department of Physics, Technion, Haifa 32000, Israel}
\date{\today}

\begin{abstract}
Axion-like particles~(ALPs) with couplings to electromagnetism have long been postulated as extensions to the Standard Model. 
String theory predicts an ``axiverse'' of many light axions, some of which may make up the dark matter in the universe and/or solve the strong CP problem. 
We propose a new experiment using superconducting radiofrequency~(SRF) cavities which is sensitive to light ALPs independent of their contribution to the cosmic dark matter density. 
Off-shell ALPs will source cubic nonlinearities in Maxwell's equations, such that if a SRF cavity is pumped at frequencies $\omega_1$ and $\omega_2$, in the presence of ALPs there will be power in modes with frequencies $2\omega_1 \pm \omega_2$. 
Our setup is similar in spirit to light-shining-through-walls~(LSW) experiments, but because the pump field itself effectively converts the ALP back to photons inside a single cavity, our sensitivity scales differently with the strength of the external fields, allowing for superior reach as compared to experiments like OSQAR while utilizing current technology.
Furthermore, a well-defined program of increasing sensitivity has a guaranteed physics result: the first observation of the Euler-Heisenberg term of low-energy QED at energies below the electron mass.
We discuss how the ALP contribution may be separated from the QED contribution by a suitable choice of pump modes and cavity geometry, and conclude by describing the ultimate sensitivity of our proposed program of experiments to ALPs.
\end{abstract}

\preprint{CERN-TH-2019-009}
\maketitle

Axions are well motivated new particles that have been proposed as a solution to the strong CP problem~\cite{Peccei:1977hh,Weinberg:1977ma,Wilczek:1977pj} (see Refs.~\cite{Sikivie:2006ni,Kim:2008hd,Hook:2018dlk} for a review).  
Additionally, string theory, predicts a plethora of light ($\ll$ eV) particles  \cite{Svrcek:2006yi}, some of which may couple to electromagnetism in a manner very similar to the axion.  
These particles have been termed axion-like particles~(ALP), and the (possibly) large number of ALPs, the ``axiverse"~\cite{Arvanitaki:2009fg}. 
One or more of these species may be excellent dark matter~(DM) candidates \cite{Preskill:1982cy,Abbott:1982af,Dine:1982ah}, and/or alleviate the hierarchy problem \cite{Graham:2015cka,Gupta:2015uea,Hook:2016mqo,Davidi:2018sii,Banerjee:2018xmn}.
In light of this strong motivation, there has been much experimental effort devoted to the axion and its cousins~\cite{Graham:2015ouw,Irastorza:2018dyq}. 

There are several general approaches for finding ALPs, roughly analogous to the multipronged approach of direct detection, indirect detection, and collider production for WIMP DM. 
If the ALP makes up the DM of the Universe, it may be detected in the laboratory by converting ALPs to electromagnetic energy (see Refs.~\cite{Zhong:2018rsr,Du:2018uak,Ouellet:2018beu} for recent experimental results) or rotating the polarization of photons \cite{DeRocco:2018jwe,Liu:2018icu}, or in radio telescopes by searching for conversion~\cite{Pshirkov:2007st,Hook:2018iia,Huang:2018lxq,Safdi:2018oeu} or decay~\cite{Caputo:2018ljp,Caputo:2018vmy} to photons in astrophysical environments.  Another approach which does not require the ALP to be DM is colloquially known as light-shining-through-walls~(LSW), where ALPs are both produced and detected in the laboratory. Such experiments are simultaneously sensitive to a wide range of ALP masses, and even multiple species of ALPs.

In this Letter, we propose a new experiment along the lines of an LSW experiment that utilizes light-by-light scattering mediated by \emph{off-shell} ALPs, with production and detection taking place in a superconducting radiofrequency~(SRF) cavity.  
An ALP $a$ is a pesudoscalar with Lagrangian
\be
	\label{eq:ALPLag}
	\mathcal{L}_a 
= 	\frac{1}{2}\partial_\mu a \partial^\mu a - \frac{1}{2}m_a^2 a^2 -\frac{1}{4}\g a F_{\mu \nu}\widetilde{F}^{\mu \nu} \, .
\ee
For processes involving photons of energy $\omega$, an ALP with mass $m_a \gg \omega$ may be integrated out, giving an effective Lagrangian \cite{Evans:2018qwy}
\be
	\label{eq:ALPLagEff}
	\mathcal{L}_{a, \rm eff} 
= 	\frac{\g^2}{32 m_a^2}(F_{\mu \nu}\widetilde{F}^{\mu \nu})^2 \, .
\ee
In other words, an off-shell ALP will induce small nonlinearities in electromagnetism. Note that this effect is \emph{local}, and does not require the ALP to propagate to another spacetime point to be converted back to photons. As we are also interested in very light ALPs, we will extend the analysis of \cite{Evans:2018qwy} to the case where $m_a \ll \omega$. In this case the nonlinear effects are nonlocal, but we will show that detection proceeds similarly to the heavy ALP case.

Famously, loop contributions from virtual electrons will also induce such nonlinearities in pure quantum electrodynamics~(QED), which are parameterized by the Euler-Heisenberg~(EH) Lagrangian~\cite{Heisenberg:1935qt,Schwinger:1951nm}. 
To lowest order in $\alpha$ and $\omega/m_e$, this is
\be
	\label{eq:LEH}
	\mathcal{L}_{\rm EH} 
= 	\frac{\alpha^2}{360 m_e^4} \left[ 4 (F_{\mu \nu}F^{\mu \nu})^2 + 7 (F_{\mu \nu}\widetilde{F}^{\mu \nu})^2 \right] \, ,
\ee
valid for $\omega \ll m_e$. Light-by-light scattering with real photons has been observed at GeV energies \cite{Burke:1997ew,Aaboud:2017bwk}, but Eq.~\eqref{eq:LEH} has never been probed with real photons at $\omega < m_e$. Thus, an experiment that is designed to look for nonlinearities induced by ALPs would, if sensitive enough, also have the guaranteed physics result of discovering light-by-light scattering at low energies for the first time ever! Crucially, the effects of ALPs and the EH Lagrangian are not exactly degenerate, as the ALP Lagrangian only contains $F_{\mu \nu}\widetilde{F}^{\mu \nu} \propto \E \cdot \B$, while the EH Lagrangian also contains $F_{\mu \nu}F^{\mu \nu} \propto \E^2 - \B^2$. Thus, the two effects may be disentangled with a suitable choice of field configurations.

Comparing Eqs.~\eqref{eq:ALPLagEff}--\eqref{eq:LEH}, we expect the ALP contribution to 4-photon processes to exceed the EH contribution when~\cite{Bernard:1997kj,Evans:2018qwy}
\be
\label{eq:EHCompare}
	\frac{\g}{m_a} 
	\gtrsim 
	\mathcal{O}(1) \times \frac{\alpha}{m_e^2} 
	\simeq 
	\frac{10^{-10} \ \GeV^{-1}}{10^{-6} \ \eV}.
\ee
The best laboratory bounds on $\g$ are from the OSQAR ~\cite{Ballou:2015cka} and PVLAS~\cite{DellaValle:2015xxa} experiments, which constrain $\g < 3.5 \times 10^{-8} \, \GeV^{-1}$ for $m_a \lesssim 10^{-4} \ \eV$.  
Surpassing these bounds with a radiofrequency experiment ($\omega \sim 10^{-6} \ \eV$) would not require sensitivity to the EH Lagrangian, since the ALP term in the Lagrangian would be much larger. Under reasonable assumptions about solar physics, the bounds from the CAST experiment~\cite{Anastassopoulos:2017ftl} constrain $\g < 6.6 \times 10^{-11} \, \GeV^{-1}$ from thermal ALPs produced in the sun.  (More stringent bounds can be obtained for $m_a \lesssim 10^{-10} \, \eV$ from the absence of photon-ALP oscillations in galactic magnetic fields \cite{Brockway:1996yr,Grifols:1996id,Payez:2014xsa,Marsh:2017yvc}.) Eq.~\eqref{eq:EHCompare} shows that an experiment which surpasses these bounds would also probe the EH contribution.  

\emph{Detection strategy.} Taking these estimates as motivation, we extend the results of Refs.~\cite{Brodin:2001zz,Eriksson:2004cz}, a novel proposal for detecting the EH Lagrangian using SRF cavities, to include the contributions from the ALP Lagrangian~\eqref{eq:ALPLag}. We will consider an SRF cavity pumped \emph{simultaneously} at two frequencies $\omega_1$ and $\omega_2$, such that ALP- or EH-induced nonlinearities will give signal photons at $\omega_s = 2\omega_1 - \omega_2$ (see Fig.~\ref{fig:schematic}). 

While it was demonstrated in~\cite{Bernard:1997kj} that light-by-light scattering experiments will likely never be sensitive to the so-called QCD axion which solves the strong-CP problem~\cite{Peccei:1977hh,Weinberg:1977ma,Wilczek:1977pj}, renewed interest in the axiverse strongly motivates a re-examination of these results for general (non-DM, non-QCD) ALPs. Indeed, multiple ALPs will all contribute to nonlinearities in electromagnetism. Because our proposed experiment is sensitive to off-shell ALPs, our signal scales as $\mathcal{N}^2_a \g^4$. By contrast, an experiment such as CAST will scale only as $\mathcal{N}_a \g^4$ whenever the ALP masses are sufficiently large such that the wavepackets corresponding to ALPs of different masses start to separate at the location of the detector, such that the amplitudes no longer add coherently. A dedicated analysis of this effect is beyond the scope of this Letter but the formalism is broadly similar to that of neutrino oscillations.


\begin{figure}[t!]
	\includegraphics[width = 0.5\textwidth]{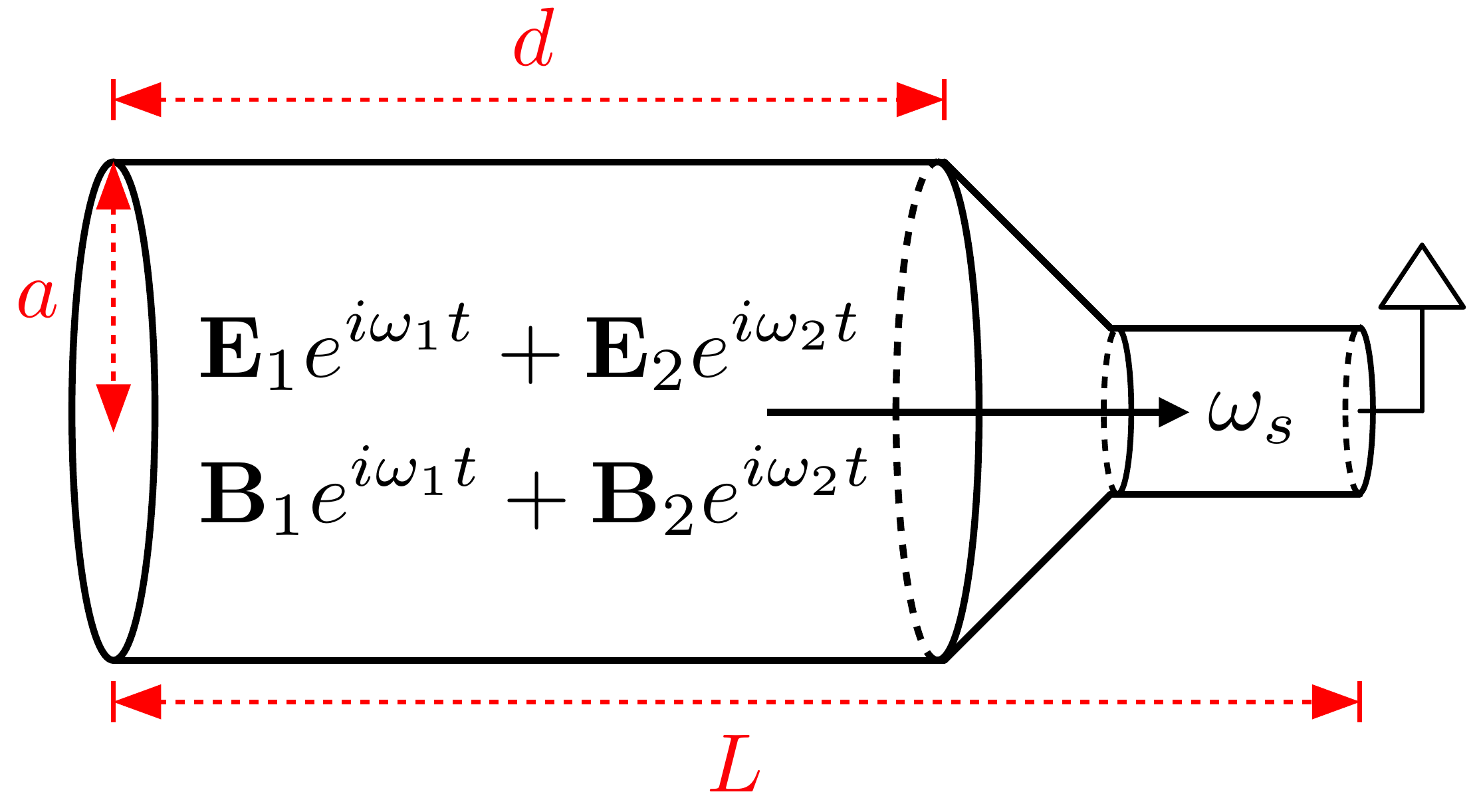}
 	\vspace{-0.7cm}
	\caption{Schematic of our proposed experiment, adapted from Ref.~\cite{Eriksson:2004cz}. Note that $\omega_s$ extends through the bulk of the cavity, but the filtering geometry suppresses the pump fields in the detection region.}
	\label{fig:schematic}
	\vspace{-0.5cm}
\end{figure}

\emph{ALP-induced cavity source terms.}
Equation~\eqref{eq:ALPLag} implies that Maxwell's equations are modified in the presence of nonzero $\g$~\cite{Sikivie:1983ip}. 
Ignoring the EH terms for now, the modified equations of motion for $\mathcal{N}_a$ ALPs with zero external charges or currents are 
\begin{align}
	\label{eq:Gauss}
	& \nabla \cdot \E  =    \B \cdot \sum_{i=1}^{\mathcal{N}_a} \g^{(i)}\nabla a_i \,, \\
	\label{eq:Ampere}
	& \nabla \times \B  = \frac{\partial \E}{\partial t}   -\E \times  \sum_{i=1}^{\mathcal{N}_a} \g^{(i)}\nabla a_i + \B   \sum_{i=1}^{\mathcal{N}_a} \g^{(i)} \frac{\partial a_i }{\partial t} \, ,  \\
	\label{eq:axioneom}
	& (\partial_t^2 - \nabla^2 + m_{a_i}^2)a_i  = \g^{(i)} \E \cdot \B \, \ \  (i = 1, \dots, \mathcal{N}_a). 
\end{align}
We will assume that the $\g^{(i)}$ are small and use classical field perturbation theory. Equation~\eqref{eq:axioneom} shows that regions of nonzero $\E \cdot \B$ will source the $a_i$ fields proportional to $\g^{(i)}$; 
Eqs.~\eqref{eq:Gauss}--\eqref{eq:Ampere} imply that $a_i$ will in turn source signal fields cubic in the cavity fields and proportional to $(\g^{(i)})^2$. If all the $\g^{(i)}$ are identical, the signal fields will be proportional to $\mathcal{N}_a\g^2$. 

Unless otherwise specified, we now restrict to the case of a single ALP, $\mathcal{N}_a = 1$. We may use the Green's function for the ALP field to write the signal fields solely in terms of the background fields. 
The appropriate Green's function is the classical retarded Green's function for the Klein-Gordon equation $G_R(\mathbf{x},t,\mathbf{x}',t')$. The solution for $a(\mathbf{x},t)$ is then
\be
	\label{eq:asolIntegral}
	a(\mathbf{x},t)  
= 	\g  \int  d^3 \mathbf{x}' dt' \, G_R(\mathbf{x},t,\mathbf{x}',t') \E(\mathbf{x}',t')  \cdot \B(\mathbf{x}',t') \, .
\ee

Suppose the cavity is pumped at resonant frequencies $\omega_1$ and $\omega_2$, with associated modes $\E_1, \B_1$ and $\E_2, \B_2$, (Fig.~\ref{fig:schematic}). The total pump field is $\E_{\rm p} = \E_1e^{i\omega_1 t} + \E_2 e^{i\omega_2 t}$, where it is understood that the physical field is the real part of the complex field and that the correct phase relationships exist between $\E$ and $\B$.  From now on, we will drop the explicit time dependence of the pump modes.

The ALP-dependent terms on the right-hand side of Eqs.~\eqref{eq:Gauss}--\eqref{eq:Ampere} can be interpreted as an effective ALP charge and current:
\be 
\label{eq:Jsol}
	\rho_a = \g \B_{\rm p} \cdot \nabla a; 
	\ \J_a = \g \! \left( \nabla a \times \E_{\rm p}  +  \B_{\rm p} \frac{\partial a }{\partial t} \right) \, . 
\ee
Note that since $a$ is quadratic in the pump fields, $\rho_a$ and $\J_a$ are cubic, with frequency components $\omega_1$, $\omega_2$, $2\omega_1 \pm \omega_2$, and $2\omega_2 \pm \omega_1$. 
If $\E_{\rm p}$ and $\B_{\rm p}$ satisfy Maxwell's equations, then $\rho_a$ and $\J_a$ satisfy the continuity equation $\partial \rho_a/\partial t + \nabla \cdot \J_a = 0$. 
Thus, using the solution for $a$ in Eq.~\eqref{eq:asolIntegral} with $\E = \E_{\rm p}$, we may treat $\J_a$ as a source for the cavity involving only the pump fields $\E_{\rm p}$ and $\B_{\rm p}$, identical in formalism to a real current source involving moving charges.

\emph{Signal strength.} 
To solve for the signal fields, we will use the general formalism of cavity Green's functions~\cite{hill2009electromagnetic}. We assume that a signal mode $\omega_s$ is a resonant mode of the cavity which matches one of the frequency components of $\J_a$, which we take to be $2\omega_1 - \omega_2$ for concreteness. Assuming a finite quality factor $Q_s$ for this mode, the ALP-sourced $\E_a$ field which develops in a cavity of volume $V$ is 
\be
	\label{eq:ESolFull}
	\E_{a} (\x) 
= 	\frac{Q_s}{\omega_s V}\hat{\E}_s(\x)\int d^3 \x' \, \hat{\E}_s(\x') \cdot \J_a(\x') \, , 
\ee
where $\hat{\E}_s$ is dimensionless with normalization $\int d^3 \x\, |\hat{\E}_{s}(\x)|^2 = V$. 

To estimate the size of the signal, we normalize the pump modes such that $\int d^3 \x\, |\E_1(\x)|^2  = \int d^3 \x\, |\E_2(\x)|^2 = E_0^2 V$, and write $\J_a = \kappa_{m_a} E_0^3 \hat{\J}_a$ where $\hat{\J}_a$ is dimensionless and $\kappa_{m_a}$ has dimension $-3$. In the two limiting cases of $m_a \gg \omega_s$ and $m_a \ll \omega_s$, we choose $\kappa_{m_a}$ to be $\kappa_\infty = \g^2 \omega_s /m_a^2$ and $\kappa_0 = \g^2/\omega_s$, respectively.
The number of photons in the signal field is
\be
	\label{eq:N3}
	N_s 
= 	\frac{1}{2 \omega_s}\int d^3 \x \, |\E_a(\x)|^2= \frac{Q_s^2  V E_0^6}{2 \omega_s^3}\kappa_{m_a}^2 K_{m_a}^2 \, ,
\ee
where we have defined the dimensionless cavity form factor
\be
	K_{m_a} 
	\equiv
  	\frac{1}{V}\left |\int d^3 \x' \, \hat{\E}_s(\x') \cdot \hat{\J}_a(\x')\right | \, .
	\label{eq:Kfac}
\ee
Note that $K$ and $\kappa$ both depend on $m_a$ through the Green's function $G_R$.

\emph{Cavity form factors: heavy and light ALPs.} 
To understand the signal strength as a function of $m_a$, we compute the cavity form factors in two limits: $K_\infty$, where $m_a \gg \omega_s$, and $K_0$, where $m_a \ll \omega_s$.

In the limit $m_a \gg \omega_s$, we have $\partial^2_t a, \nabla^2 a \ll m_a^2 a$, so we can ``integrate out'' the ALP by by solving algebraically for $a$ in terms of the pump fields. This gives
%
\be
	\label{eq:Jheavy}
	\J_\infty 
\!=\! 	\frac{\g^2}{m_a^2}
	\!\left(  
	\nabla (\E_{\rm p} \cdot \B_{\rm p}) \times \E_{\rm p} 
	+ \B_{\rm p} \frac{\partial}{\partial t} \left(\E_{\rm p} \cdot \B_{\rm p}\right) \right)  .
\ee

In the limit $m_a \to 0$, the ALP Green's function is identical to the retarded Green's function familiar from electromagnetism:
\be
\label{eq:a0sol}
	a_0(\x, t) 
= 	\frac{\g}{4\pi} \int  \frac{d^3 \mathbf{x}'}{|\x - \x'|} \E_{\rm p}(\mathbf{x'}, t_R) \cdot \B_{\rm p}(\mathbf{x'}, t_R) \, ,
\ee
where $t_R = t-|\x - \x'|$ is the retarded time. 
In this case, $a$ responds nonlocally to changes in $\E_{\rm p}$ and $\B_{\rm p}$, with a time delay given by $t_R$. 
Since the ALP-mediated current  $\J_0$, which may be computed from Eq.~\eqref{eq:a0sol} using Eq.~\eqref{eq:Jsol}, is also nonlocal, there is no simple expression in terms of the pump fields.

As an example, consider a right cylindrical cavity with $\omega_1 = {\rm TE}_{011}$, $\omega_2 = {\rm TM}_{010}$, and $\omega_s = 2\omega_1 - \omega_2 = {\rm TM}_{020}$ (with mode labeling conventions following \cite{hill2009electromagnetic}), satisfied for a cavity of radius $a$ and height $d = 3.112 a$. We find $K_\infty = 0.18$ and $K_0 = 0.24$ (see Supplementary Material (SM) for details), where the latter result assumes that both $\sin(\omega_s t)$ and $\cos(\omega_s t)$ components of the signal can be added in quadrature, appropriate for photon counting at the standard quantum limit. This example demonstrates that the cavity form factor can be relatively insensitive to $m_a$, and thus a single cavity can be used to probe a broad range of ALP masses. This is a strong advantage over traditional resonant searches for ALP dark matter~\cite{Du:2018uak,Zhong:2018rsr,Sikivie:2013laa,JacksonKimball:2017elr}, which require careful tuning to match a resonance (e.g.\ a cavity mode or a Larmor frequency) to $m_a$. By contrast, once $\omega_s = 2\omega_1 - \omega_2$ is accomplished by tuning the cavity geometry, no further tuning is required to set limits on $\g$ for any $m_a$.

\emph{Expected sensitivity to ALPs.} 
To estimate the sensitivities in the light and heavy mass limits, we compute the expected number of signal photons $N_s$. From Eq.~\eqref{eq:N3} we have
\be
	\label{eq:finalnumber}
	N_{s, 0} =  \frac{Q_s^2 V E_0^6 }{2\omega_s} \frac{ \g^4 }{ \omega_s^4} K_{0}^2 \, , \ 
	N_{s, \infty}  = \frac{Q_s^2 V E_0^6 }{2\omega_s} \frac{ \g^4 }{ m_a^4} K_{\infty}^2 \, .
\ee

To measure the signal, we imagine a filtering geometry as suggested in Ref.~\cite{Eriksson:2004cz} (and shown in Fig.~\ref{fig:schematic}) where at some point in the geometry the pump fields are exponentially suppressed compared to the signal field, which is possible as long as $\omega_s > \omega_{1,2}$.  
At this location, the signal can be measured without contamination from the pump modes (though we note that this is more of a practical concern than an irreducible background, since the pump modes are at a different frequency than the signal). We estimate the signal-to-noise ratio (SNR) with the Dicke radiometer equation, neglecting any information about the field phase:
\be
	\SNR = \frac{P_s}{T} \sqrt{\frac{t}{B}} \approx \frac{N_s}{N_{\rm th}} \frac{1}{2 L Q_s} \sqrt{\frac{t}{B}} \, ,
\ee
where $P_s$ is the signal power, $t$ is the total measurement time, $B$ is the signal bandwidth, $L$ is the length of the cavity, and $N_{\rm th} = T/\omega_s$ is the number of thermal photons at the signal frequency (valid for temperatures $T \gg \omega_s$, and assuming thermal noise dominates). A detailed sensitivity calculation exploiting our knowledge of the pump field phases, perhaps using phase-sensitive amplifiers instead of photon counting, will be presented in a future work.

Our expected sensitivity to $\g$ is then
\be
	\label{eq:scaling}
	\g^{\rm lim.} 
= \left(\frac{4 T L}{Q_s V E_0^6} \sqrt{\frac{B}{t}} \, \SNR \right)^{1/4} \times	\begin{cases}
		K_0^{-1/2} \omega_s ,  & m_a \ll \omega_s\\
		K_\infty^{-1/2} m_a,  & m_a \gg \omega_s \, .
	\end{cases}
\ee
In an actual experimental implementation, a cavity should be designed specifically to maximize the figure of merit in Eq.~\eqref{eq:scaling} while minimizing issues such as multipacting, dark currents, field emission, intermodulation in the feed lines, and surface nonlinearities \cite{padamsee2001science,padamsee2009rf}.

For a fixed choice of modes and cavity size (hence fixed $\omega_s$ and $V$), the reach is constant at small $m_a$ and degrades linearly at large $m_a$. 
There is in principle some dependence of $K_{m_a}$ on $m_a$, but with a suitable choice of modes this dependence is extremely mild. 
For a large number $\mathcal{N}_a$ of light ALPs, all with $m_a \ll \omega_s$, Eq.~\eqref{eq:scaling} should be interpreted as a limit on $\sqrt{\sum_{i=1}^{\mathcal{N}_a} \left( \g^{(i)} \right)^2 }$.



\begin{figure}[t]
\includegraphics[width = 0.5\textwidth]{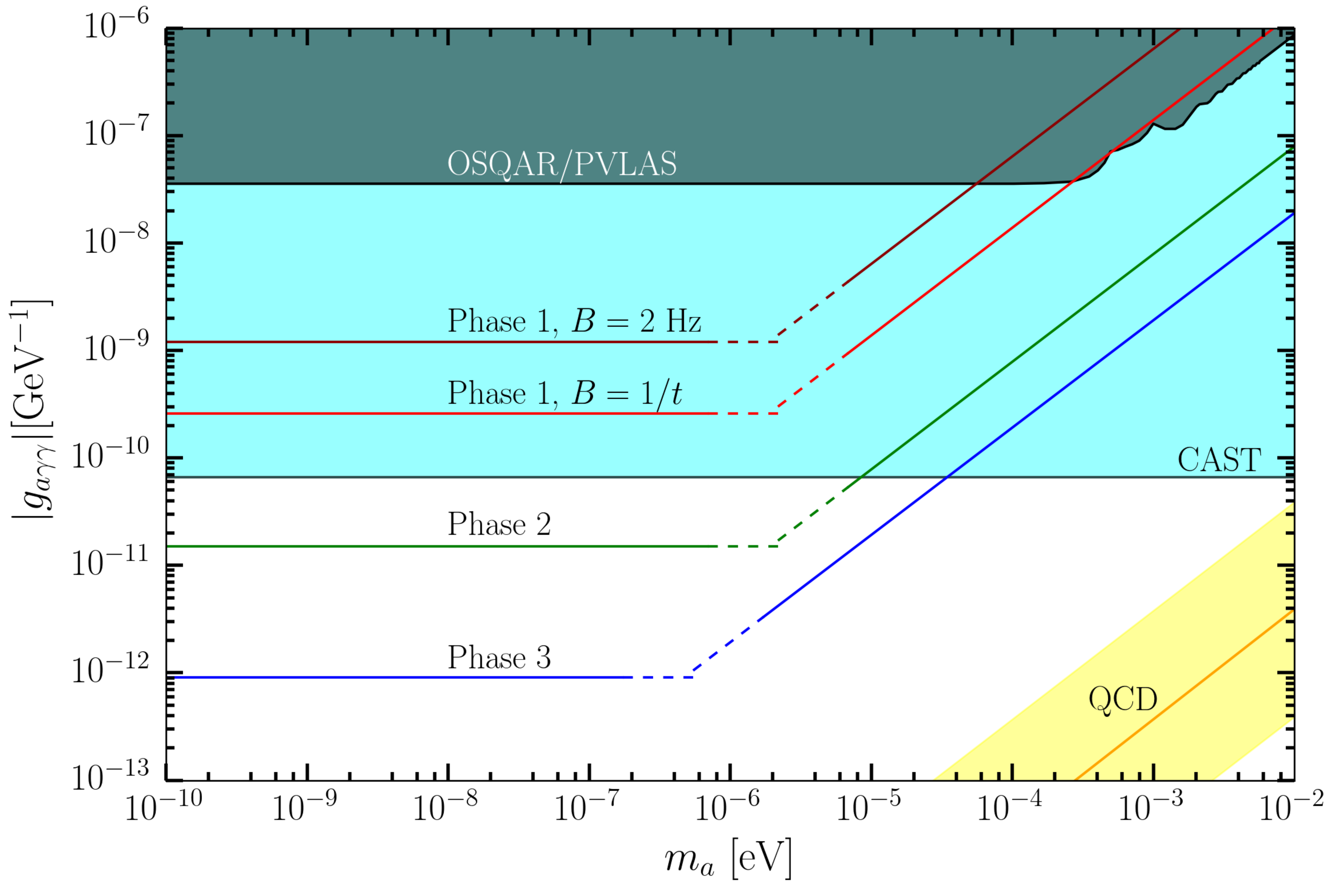}
\caption{Projected sensitivities of our proposal (approximations for $m_a \sim \omega_s$ shown in dashed lines), along with existing constraints \cite{Ballou:2015cka,DellaValle:2015xxa,Anastassopoulos:2017ftl}. See text for details.}
\vspace{-0.5cm}
\label{fig:moneyplot}
\end{figure}

\emph{Phase 1: Conservative projected reach.} 
We envision our experiment progressing in three stages, each building on current technology. For Phase 1, we take the following parameters: cavity temperature $T=1.5$\,K; a right cylindrical cavity with $a = 0.5$ m and the ${\rm TE}_{011}/{\rm TM}_{010}/{\rm TM}_{020}$ mode combination, giving $d = 1.56$ m, $f_s  = \omega_s/(2\pi) = 527 \, {\rm MHz}$ and $V = 1.23 \, {\rm m}^3$; $K_0 = 0.24$ as calculated above; pump field strength $E_0 = 45 \, {\rm MV}/{\rm m}$; and a cavity bandwidth of $f_s/Q_s = 2$ Hz, corresponding to $Q_s = 2.6 \times 10^{8}$. This $Q$ is much smaller than what typical high-performance SRF cavities can achieve, but a wide cavity response function for Phase 1 allows the frequency-matching condition $\omega_s = 2\omega_1 - \omega_2$ to be approximately satisfied even if vibrational distortions shift $\omega_s$ by $\mathcal{O}({\rm Hz})$. Note also that our mode combination satisfies $\omega_s > \omega_1, \omega_2$, making it amenable to filtering; to model this, we assume the total cavity length is twice the cavity height, $L = 3.11$ m. 

At 1.5\,K, the thermal noise in the signal mode is $N_{\rm th} = kT/\omega_s \simeq 60$ photons. A lower operating temperature would be desirable, but the cooling power requirements are substantial: assuming that the pump modes have $Q_{1,2} = 10^{12}$, characteristic of the best $Q$ achieved in SRF cavities \cite{doi:10.1063/1.4903808,Romanenko:2018nut}, the cavity lifetime for the pump modes is $\tau_{1,2} = Q_{1,2}/f_{1,2} \sim 2600\,{\rm s}$ and the power dissipated is $\mathcal{O}(10 \, {\rm W})$. Dilution refrigerators, which have a cooling capacity of $\mathcal{O}$(mW), are not sufficient, and the cavity must operate at liquid helium temperatures. 

We first consider the case where the injected pump bandwidth is comparable to the cavity bandwidth, $B = 2\,$Hz. 
For light ALPs, the OSQAR bound can be surpassed by nearly an order of magnitude in a measurement of a single cavity lifetime of $\tau_s = Q_s/f_s = 0.5$\,s.
Integrating the signal over a time $t \sim $ 1 day, we can obtain a Phase 1 reach of $g^{\rm lim.}_{a\gamma\gamma, 0} = 1.2 \times 10^{-9} \, \GeV^{-1}$. For $m_a \gg \omega_s$, we can get the limits by using $g^{\rm lim.}_{a\gamma\gamma,\infty} = (m_a/\omega_s)\sqrt{K_0/K_\infty} \, g^{\rm lim.}_{a\gamma\gamma,0}$. 

One could also pump the cavity with a bandwidth narrower than the cavity itself, for example by locking the pump tones to an atomic clock.  Taking $B = 1/t$, the narrowest allowed bandwidth for a given measurement time $t$, a bound of $g^{\rm lim.}_{a\gamma\gamma, 0} = 2.6 \times 10^{-10} \, \GeV^{-1}$ could be reached in a day. In the case $B = 1/t$, the SNR scales linearly with time, so the limit on $\g$ scales as $t^{1/4}$. The two bandwidth choices for Phase 1 are shown in Fig.~\ref{fig:moneyplot}; we have not explicitly calculated the reach for $m_a \sim \omega_s$ (shown as dashed lines), but we expect the light and heavy mass limits to be excellent approximations away from this region.


\emph{Phase 2: Detecting the Euler-Heisenberg contribution.} 
As we have discussed, there is an irreducible contribution to cubic nonlinearities in Maxwell's equations from the EH Lagrangian, see Eq.~\eqref{eq:LEH}.
The effective EH charge and current are~\cite{Soljacic:2000zz,Brodin:2001zz}
\begin{align}
\label{eq:JEH}
	\rho_{\rm EH} \!=\! - \frac{4\alpha^2}{45m_e^4} \nabla \cdot \Pol \, ; \ 
	\J_{\rm EH} \!=\! \frac{4\alpha^2}{45m_e^4} \! \left( \nabla\! \times\! \M \!+\! \frac{\partial \Pol}{dt} \right)  ,
\end{align}
with $\Pol = 7(\E\cdot \B)\B + 2 (\E^2 - \B^2)\E \,$ and $\M = 7(\E\cdot \B)\E - 2 (\E^2 - \B^2)\B\,$.
The number of photons from the EH signal can be estimated similarly to the ALP case.

For Phase 2, we assume the cylindrical cavity geometry from Phase~1, but with $Q_s = 10^{12}$. Indeed, in tuned SQUID magnetometers, a feedback circuit may be used to broaden the bandwidth without sacrificing $Q$ \cite{2007JMagR.186..182M}; such a scheme may be possible here. We find that 
\begin{align}
	\label{eq:NEH}
	N_{\rm EH} 
=	\frac{Q_s^2 V E_0^6 }{2\omega_s^3 }\kappa_{\rm EH}^2 K_{\rm EH}^2 \approx 3.6 \, 
\end{align}
with $\kappa_{\rm EH} = 4 \alpha^2 \omega_s /(45m_e^4)$ and $K_{\rm EH}=\frac{1}{V}\int d^3 \x' \hat{\J}_{\rm EH} \cdot \hat{\E}_s =0.18$, with $\hat{\J}_{\rm EH}$ defined analogously to $\hat{\J}_{a}$. This signal strength is roughly consistent with Ref.~\cite{Eriksson:2004cz} given our different choices of parameters and modes. Therefore, assuming $B=1/t$, the EH signal can be detected within 20 days of running. The corresponding sensitivity to light ALPs for the same integration time is $g^{\rm lim.}_{a\gamma\gamma, 0} = 1.6 \times 10^{-11} \ \GeV^{-1}$; this is shown in Fig.~\ref{fig:moneyplot}. This would surpass the CAST bound of $\g = 6.6 \times 10^{-11} \, \GeV^{-1}$ and would also be competitive with recent proposals to search for ALP DM at low masses such as ABRACADABRA~\cite{Kahn:2016aff,Ouellet:2018beu}. In some models, an ALP with these couplings could also be the QCD axion \cite{Farina:2016tgd,Agrawal:2017cmd}. If a positive signal were detected, the ALP nature of the signal could be verified using a second cavity with different mode combinations. If the ALP is heavier than $\omega_s$, the combination of the two measurements would suffice to determine both $\g$ and $m_a$.

Naively, the sensitivity of this proposal to probe ALPs becomes limited when $N_{\rm EH}\sim N_s$. In principle, one can search for the ALP signal on the top of the thermal and EH backgrounds, but as with the ``neutrino floor'' in WIMP direct detection experiments, the SNR will grow much slower than $t^{1/4}$. However, with a slightly different mode choice, the EH contribution can be removed, leaving behind only the ALP signal. The idea is to pump an additional mode degenerate with $\omega_1$ but with a different field configuration. By tuning the three different pump amplitudes, we can arrange to have $K_{\rm EH}=0$ with $K_{m_a}\ne0$. For these special pump amplitudes, the EH contribution to light-by-light scattering vanishes at \emph{amplitude} level, and there is no interference with the ALP amplitude.  We give a proof of principle demonstration of this idea in the SM.

\emph{Phase 3: Probing the axiverse.} 
As the optimistic endpoint of this proposed program of experiments, consider a large cylindrical cavity with $a = 2$\,m and $d = 6.22$\,m, giving $f_s =  132$\,MHz, with the same mode combinations as considered in Phases 1 and 2 and $Q_s = 10^{12}$. We suppose a cavity geometry can be developed which permits $K_0 \sim 0.24$ with the EH contribution tuned away to sufficient precision as described in Phase 2, and a compact filtering geometry with length $L = 10$\,m. Assuming the same pump strength as Phases 1 and 2, and integrating for a total time $t = $ 1\,year with $B = 1/t$, we find from Eq.~\eqref{eq:scaling} a maximum sensitivity at low masses of $g^{\rm lim.}_{a\gamma\gamma,0} \sim 9.1 \times 10^{-13} \ \GeV^{-1}$, shown in Fig.~\ref{fig:moneyplot}.

Revisiting the axiverse scenario, suppose that $\mathcal{N}_a$ ALPs all had decay constants $f_a$ at the string scale, which we conservatively take to be the renormalized Planck scale, $10^{18} \, \GeV/\sqrt{\mathcal{N}_a}$.  
These string ALPs would have photon couplings of $ \g = \alpha/f_a \sim \sqrt{\mathcal{N}_a} 10^{-20} \, \GeV^{-1}$. Our Phase 3 would be sensitive to $\g \sqrt{\mathcal{N}_a} \sim 10^{-20} \ \GeV^{-1} \times \mathcal{N}_a$, which could bound the number of string-scale ALPs with masses less than $10^{-6} \, \eV$ by $\mathcal{N}_a \lesssim 10^{8}$. While this is still (much) larger than typical expectations from string theory, one could still imagine placing constraints on particular compactification geometries which contain large numbers of nontrivial cycles  \cite{Douglas:2006es}, allowing low-energy SRF cavity experiments to offer a fascinating probe into the ultra-high-energy regime of quantum gravity and the landscape of string theory vacua.

{\it Acknowledgments. ---} 
AH, YK, and YS thank Ben Safdi and the University of Michigan Slack channel for facilitating discussion in the early stages of this work. YK thanks Prateek Agrawal, M.C. David Marsh, and the participants of the workshop ``Axions in Stockholm -- Reloaded'' for discussions about the axiverse, and James Halverson and Cody Long for discussions about axions in string compactifications. We thank Daniel Bowring, Fritz Caspers, Aaron Chou, Anna Grassellino, Roni Harnik, Kent Irwin, Akira Miyazaki, Jonathan Ouellet, Sam Posen, Alexander Romanenko, and Slava Yakovlev for enlightening discussions regarding cavity design and photon readout. We thank Junwu Huang, Gilad Perez and Jesse Thaler for helpful comments, and Lindley Winslow for support in the early stages of this project.  AH is supported in part by the NSF under Grant No. PHY-1620074 and by the Maryland Center for Fundamental Physics (MCFP).  The work of ZB is supported by the National Science Foundation under grant number NSF PHY-1806440.  This work was supported in part by the Kavli Institute for Cosmological Physics at the University of Chicago through an endowment from the Kavli Foundation and its founder Fred Kavli.

\bibliography{AxionBib.bib}

\clearpage
\newpage
\maketitle
\onecolumngrid
\begin{center}
\textbf{\large Probing ALPs and the Axiverse with Superconducting Radiofrequency Cavities} \\ 
\vspace{0.05in}
{ \it \large Supplemental Material}\\ 
\vspace{0.05in}
{}
{ Zachary Bogorad, Anson Hook, Yonatan Kahn, Yotam Soreq}

\end{center}
\setcounter{equation}{0}
\setcounter{figure}{0}
\setcounter{table}{0}
\setcounter{section}{1}
\setcounter{page}{1}
\renewcommand{\theequation}{S\arabic{equation}}
\renewcommand{\thefigure}{S\arabic{figure}}
\renewcommand{\thetable}{S\arabic{table}}
\newcommand\ptwiddle[1]{\mathord{\mathop{#1}\limits^{\scriptscriptstyle(\sim)}}}

In this Supplemental Material, we compare our proposal to LSW experiments as well as multimode pumping for axion DM; give further details about the example mode choices we have used to calculate the cavity form factors; discuss the choice of modes for optimizing our reach for both light and heavy ALPs; and give an example of degenerate mode tuning which can be used to eliminate the EH contribution to a Phase 3 ALP search.

\section*{Comparison to other experiments}

The general setup of an LSW experiment is as follows: a laser passes through a large magnetic field $B_{\rm prod.}$, some photons convert to ALPs, a wall blocks the remaining photons but not the ALPs, and after the wall another large magnetic field $B_{\rm det.}$ converts the ALPs back into detectable photons.  
Let us briefly compare the parametrics of our proposal to an LSW experiment. In the heavy ALP case, $m_a \ll \omega$, the number of signal photons per number $N_i$ of input photons scales like
\be
	\left. \frac{N_s}{N_i} \right|_{\rm LSW} \mkern-30mu \sim \g^4 B_{\rm prod.}^2 B_{\rm det.}^2 L^4  , \ 
	\left. \frac{N_s}{N_i} \right|_{\text{cavity}} \mkern-30mu \sim Q_s^2 \g^4 E_0^4 L^4,
\ee
where $L$ is the typical size of the experiment (for our setup, we assume $L \sim 1/\omega_s$). Our experiment scales similarly to an LSW experiment except that our final number of photons has been enhanced by $Q_s^2$ due to the cavity, and there is only a single field region rather than separate production and detection regions; instead of $B_{\rm prod.}$ and $B_{\rm det.}$, the input oscillating field $E_0$ does the conversion.  A static $B$ field can be made about 40 times larger than an oscillating $B$ field, but this deficit is more than made up for by the large $Q$ factor of SRF cavities, which can reach $10^{12}$ \cite{doi:10.1063/1.4903808,Romanenko:2018nut}. 

It is worth noting that the scaling of an LSW experiment utilizing a cavity, e.g. ALPS-II, would also be enhanced by $Q$, but such cavities could not be made superconducting at $B$-fields above a few T, and the largest $Q$ achievable in copper cavities is about $10^6$.  Our cavity experiment would still have superior sensitivity to such an LSW experiment, unless the ALP were both produced and detected in high-$Q$ SRF cavities, where the larger $Q$ compensates for the smaller $B$~\cite{DarkSRFRoni,DarkSRFAnna}. 
Furthermore, in our setup, the detection frequency is not a harmonic of the input frequency, which may help reduce backgrounds. 
In the large mass limit of an LSW experiment, $N_s/N_i \sim 1/m_a^8$, until $\omega \sim m_a$ and ALP-photon conversions cannot occur.  
In the large mass limit of our proposal, $N_s/N_i \sim 1/m_a^4$, giving superior reach at large masses.

While our proposal exploits multimode pumping for virtual ALP detection, multimode pumping and heterodyne readout has also been considered for the case where the ALP comprises the cosmic DM abundance \cite{Sikivie:2010fa,Goryachev:2018vjt,Goryachev:2018woh}. However, for axions heavier than $\sim 10^{-16} \ \eV$, this fails to improve on traditional searches using static $B$-fields because of the DM velocity dispersion. Note that in a search for DM axions using a pumped cavity, the axion DM field may be considered as a pump field at frequency $\omega = m_a$, with intrinsic bandwidth set by the velocity dispersion, $B \sim v_{\rm DM}^2 m_a \sim 10^{-6} m_a \sim \kHz$ for $m_a \sim 10^{-6} \ \eV$. This limits the scaling of the SNR with time when $B > 1/t$ and the maximum $Q_s$ given the necessity to scan \cite{Sikivie:2010fa}, an effect not accounted for in Refs.~\cite{Goryachev:2018vjt,Goryachev:2018woh}.

\section*{Mode functions and signal currents}

\begin{figure}[t]
	\includegraphics[width = 0.5\textwidth]{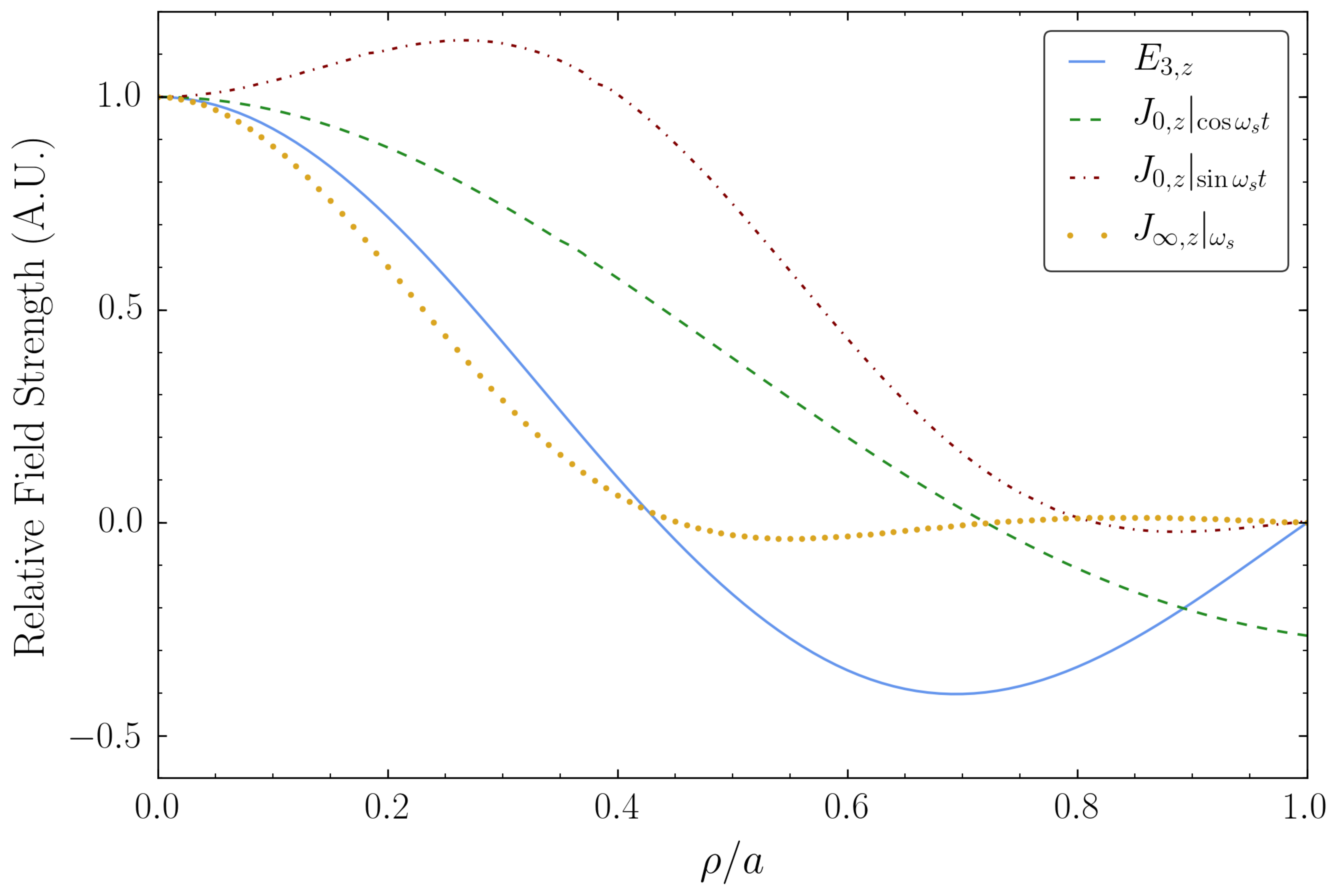}
 	\vspace{-0.3cm}
	\caption{Signal mode $E_{s,z}$, along with ALP-induced effective currents $J_{\infty,z}|_{\omega_s}$ and $ J_{0,z}|_{\omega_s}$ and EH effective current $J_{{\rm EH},z}|_{\omega_s}$ for the TE$_{011}$/TM$_{010}$/TM$_{020}$ mode combination. There is no $\phi$ dependence in either the signal mode or the effective currents; the mode profiles are evaluated at $z = d/2$, normalized to 1 at $\rho = 0$, and plotted as a function of the remaining variable $\rho$. The similar profile of the signal mode and the effective currents leads to a large form factor $K$ for this mode choice.}
	\label{fig:JCompare}
	\vspace{-0.5cm}
\end{figure}

Here we explicitly calculate the axion current and overlap for the mode choices $\omega_1 = $ TE$_{011}$, $\omega_2 = $ TM$_{010}$, and $\omega_s = 2 \omega_1 - \omega_2 = $ TM$_{020}$ in a cylindrical cavity of height $d$ and radius $a$, using mode conventions from Ref.~\cite{hill2009electromagnetic}. The (un-normalized) $E$-field of the signal mode only has a $z$-component:
\be
\E_s = E_0 J_0\left(\frac{x_{02} \, \rho}{a}\right)\hat{\mathbf{z}},
\ee
where $J_0$ is the Bessel function of order 0 and $x_{02}$ is its second zero. Thus the form factor integrand \eqref{eq:Kfac} only receives a contribution from $J_z$. The (un-normalized) pump fields are
\begin{align}
\E_1 & = E_0  \, \omega_1 \frac{x_{01}'}{a} \left(J_0'(\rho) \sin \frac{\pi z}{d}\hat{\bm{\phi}}\right)\sin (\omega_1 t) \label{eq:E1}\\
\E_2 & = E_0 \,  \frac{x_{01}^2}{a^2} J_0(\rho) \hat{\mathbf{z}} \cos (\omega_2 t) \label{eq:E2}\\
\B_1 & = E_0 \, \frac{(x_{01}')^2}{a^2} \left(\frac{\pi}{d}\frac{x'_{01}}{a} J_0'(\rho) \cos \frac{\pi z}{d}\hat{\bm{\rho}} + J_0(\rho) \sin \frac{\pi z}{d}\hat{\mathbf{z}}\right)\cos(\omega_1 t) \label{eq:B1}\\
\B_2 & = -E_0 \,  \omega_2 \frac{x_{01}}{a} J_0'(\rho)\hat{\bm{\phi}}\sin(\omega_2 t) \label{eq:B2}
\end{align}
where $x_{01}$ is the first zero of $J_0$ and $x'_{01}$ is the first zero of $J'_0$. The component of $\J$ with frequency $2\omega_1 - \omega_2$ will contain two mode 1 fields and one mode 2 field, i.e. terms like $\E_1^2 \B_2$. 

For the heavy mass case $m_a \to \infty$, inspecting Eq.~(\ref{eq:Jheavy}) and keeping track of the time dependence, we have that the component of $J_a$ quadratic in mode 1 and linear in mode 2 is
\be
J^{(112)}_{\infty,z} = E_{1,\phi}s_1 \frac{\partial}{\partial \rho}\bigg(E_{1,\phi}B_{2,\phi} s_1 s_2 + E_{2,z}B_{1,z}c_1 c_2\bigg)
+ B_{1,z}c_1\bigg(E_{1,\phi}B_{2,\phi}(\omega_1 c_1 s_2 + \omega_2 s_1 c_2) - E_{2,z}B_{1,z}(\omega_1 s_1 c_2 + \omega_2 c_1 s_2)\bigg)
\ee
where $s_{1,2} = \sin (\omega_{1,2} t)$ and $c_{1,2} = \cos(\omega_{1,2} t)$.

At this point, $J^{(112)}_{\infty,z}$ has frequency components $\omega_2$, $2\omega_1 - \omega_2$, and $2\omega_1 + \omega_2$. We now wish to isolate the frequency component at $\omega_s = 2\omega_1 - \omega_2$. To do this, we note that terms appear such as
\be
\sin^2(\omega_1 t)\sin(\omega_2 t) = \frac{1}{2}\sin(\omega_2 t) + \frac{1}{4}\sin((2\omega_1 - \omega_2)t) - \frac{1}{4}\sin((2\omega_1 + \omega_2)t),
\ee
so to isolate the desired frequency component, we make the replacement
\be
s_1^2 s_1 \to \frac{1}{4}.
\ee
Similarly, for the other two terms we have
\be
s_1 c_1 c_2  \to \frac{1}{4}, \qquad
c_1^2 s_2  \to -\frac{1}{4},
\ee
where in all three cases only the $\sin(\omega_s t)$ phase component appears (i.e. there is no $\cos(\omega_s t)$ term). Thus the component of $J^{(112)}_{a,z}$ oscillating at the signal frequency is
\be
\label{eq:J112ex}
 J_{\infty,z}|_{\omega_s} = \frac{1}{4} \left \{ E_{1,\phi}\frac{\partial}{\partial \rho}\bigg(E_{1,\phi}B_{2,\phi} + E_{2,z}B_{1,z}\bigg) 
+ B_{1,z}(\omega_2 - \omega_1)\bigg(E_{1,\phi}B_{2,\phi} + E_{2,z}B_{1,z}\bigg) \right \}\sin (\omega_s t).
\ee

Plugging \eqref{eq:E1}--\eqref{eq:B2} into Eq.~(\ref{eq:J112ex}), we see that the $z$-dependence of all terms is $\sin^2\left(\frac{\pi z}{d}\right)$, and there is no $\phi$ dependence. Evaluating Eq.~(\ref{eq:J112ex}) with $\omega_1 = 3.96245/a$ and $\omega_2 = 2.40483/a$ which satisfies the frequency-matching condition for the third mode at $z = d/2$, we obtain  $J_{\infty,z}(\rho)|_{\omega_s}$, which is plotted in Fig.~\ref{fig:JCompare}. For comparison, we also plot $E_{3,z}(\rho)$ with both profiles normalized to 1 at $\rho = 0$, showing that the shape of these functions is fairly similar and we expect a large overlap. 

As noted in the main text, the current $\mathbf{J}_0$ in the light mass case is nonlocal, so there is no simple analytic expression in terms of the pump fields. Nonetheless, we can evaluate the spatial integrals in Eq.~\eqref{eq:Kfac} numerically, and isolate the frequency components as described above. Unlike the heavy mass case, both phase components $\sin(\omega_s t)$ and $\cos(\omega_s t)$ are present. The two phase components of $\J_{0,z}$ are also shown in Fig.~\ref{fig:JCompare}; note that the component which is in phase with the pump modes vanishes at the boundary $\rho = a$, while the other phase component does not. Similar to the axion contribution, we can also calculate the EH effective current $J_{{\rm EH},z}|_{\omega_s}$ from \eqref{eq:JEH}, which is also shown in Fig.~\ref{fig:JCompare}.

\section*{Characteristics of light and heavy form factors}
In order to test the feasibility of our proposed method, we searched through a number of cylindrical cavity mode combinations and calculated the expected coupling $K_\infty$ for each. Since there are, in principle, infinitely many possible mode combinations, we restricted to the six smallest non-trivial mode numbers for each field and mode number. We also took advantage of three selection rules for modes TE$_{npq}$ and TM$_{npq}$:
\begin{itemize}
    \item Either $\pm 2n_1 \pm n_2 = n_s$ (including all sign combinations) or $n_2 = n_s$.
    \item Either $\pm 2q_1 \pm q_2 = q_s$ (including all sign combinations) or $q_2 = q_s$.
    \item If $\omega_2$ and $\omega_s$ are both TE modes or both TM modes, then $\E_s$ and $\B_s$ must have the same $\cos{n\phi}$ dependence as $\E_{1,2}$ and $\B_{1,2}$, rather than $\sin{n\phi}$.
\end{itemize}

We found several modes with $K_\infty$ in the range $0.1$--$0.2$. We then chose five of these with generally smaller mode numbers and calculated $K_0$ for each in order to test whether the same cavity dimensions would allow for effective searches of both high- and low-mass ALPs. As noted above, because $K_0$ contains both phase components $\sin(\omega_s t)$ and $\cos(\omega_s t)$, we compute $K_0$ by summing the form factors in quadrature for the two phase components. As described in the main text, this is appropriate for photon counting at the standard quantum limit, but in future work we will explore the benefits of phase-sensitive amplifiers, in which case one quadrature may dominate. The values of $K_\infty$ and $K_0$ for each of these five modes are given in Table \ref{tab:KInftyAnd0}.

\begin{table}[t]
\centering
\begin{tabular}{|c|c|c||c|c|}
	\hline $\omega_1$ & $\omega_2$ & $\omega_s$ & $K_\infty$ & $K_0$ \\
	\hline TE$_{011}$ & TM$_{010}$ & TM$_{020}$ & 0.18 & 0.24 \\
	\hline TE$_{011}$  & TM$_{011}$ & TM$_{013}$ & 0.12 & 0.080 \\
	\hline TE$_{011}$  & TM$_{012}$ & TM$_{014}$ & 0.13 & 0.075 \\
	\hline TE$_{011}$  & TM$_{030}$ & TM$_{050}$ & 0.11 & 0.079 \\
	\hline TE$_{011}$  & TM$_{040}$ & TM$_{060}$ & 0.12 & 0.062 \\
	\hline
\end{tabular}
\caption{The values of $K_\infty$ and $K_0$ for five mode combinations with relatively large values of $K_\infty$ and small mode numbers. Note that, for all five mode combinations, $K_0$ is within a factor of 2 of $K_\infty$, making it reasonable to use the same cavity to search for both high- and low-mass ALPs.}
\label{tab:KInftyAnd0}
\end{table}

We conclude that a precise choice of modes is not necessary for achieving a large cavity form factor for both heavy and light ALPs, though we find that the largest form factors come from $\omega_1 = {\rm TE}_{011}$ and $\omega_2$ and $\omega_s$ being TM modes. These general properties are easy to reproduce in an elliptical cavity, which will likely be the basis for a realistic design.

\section*{An Euler-Heisenberg example}

Here, we demonstrate that the Euler-Heisenberg contribution to signal photon production can be canceled with a careful choice of pump modes and relative field strengths.  We consider a rectangular cavity of dimensions $a\times b \times c$ and the limit of $m_a\gg \omega_s$. 
The three pump modes (labeled 1, $1'$, and 2) are ${\rm TE}_{221}/{\rm TM}_{221}/{\rm TM}_{121}$ and the signal mode is ${\rm TM}_{163}\,$. The matching condition $\omega_s = 2\omega_1 - \omega_2$ is satisfied for $b=4a$ and $c=1.22 a\,$. The total pump field is $\E_{\rm p} = r_1 \E_1 + r_{1'} \E_{1'} + r_2 \E_2$ (where $r_1$, $r_{1'}$, and $r_2$ are dimensionless) and similarly for $\B_{\rm p}$. We find 
\begin{align}
	K_{\infty} 
=& 	0.047 r_2  (r_1^2 - 0.18 r_{1'}^2) \, ,  \nonumber\\
	K_{\rm EH} 
=& 	0.059 r_2  (r_1^2 - 8.24r_{1'}^2) \, .
\end{align}
Therefore, for $r_{1'} = 0.35 r_{1}$ we get $K_{\rm EH}=0$ and $K_{\infty} = 0.046 r_1^2 r_2\,$.

In the above example, we chose a rectangular cavity for simplicity because TE and TM modes with the same mode numbers are automatically degenerate, and because there is an additional free parameter in the cavity geometry which permits the correct configuration of form factors. We note that this idea can also be implemented with an elliptical cavity, which avoids the large field gradients present at the corners of rectangular cavities, and which may also be used for the Phase 1 search described in the main text.

\end{document}